# Energy calibration of bulk events in the BULLKID detector

**M. Folcarelli**[1,2,a], **D. Delicato**[1,2,7], **A. Acevedo-Rentería**[5], **L. E. Ardila-Perez**[9], **L. Bandiera**[6], **M. Calvo**[7], **M. Cappelli**[1,2], **R. Caravita**[10], **F. Carillo**[4], **U. Chowdhury**[7], **D. Crovo**[9], **A. Cruciani**[2], **A. D'Addabbo**[8], **M. De Lucia**[3,4], **G. Del Castello**[2], **M. del Gallo Roccagiovine**[1,2], **F. Ferraro**[8], **S. Fu**[8], **R. Gartmann**[9], **M. Grassi**[4], **V. Guidi**[6,11], **D. Helis**[8], **T. Lari**[3,4], **L. Malagutti**[6], **A. Mazzolari**[6,11], **A. Monfardini**[7], **T. Muscheid**[9], **D. Nicolò**[3,4], **F. Paolucci**[3,4], **D. Pasciuto**[2], **L. Pesce**[1,2], **C. Puglia**[4], **D. Quaranta**[1,2], **C. M. A. Roda**[3,4], **S. Roddaro**[3,4], **M. Romagnoni**[6], **G. Signorelli**[3,4], **F. Simon**[9], **A. Tartari**[4], **E. Vázquez-Jáuregui**[5], **M. Vignati**[1,2], **K. Zhao**[8,12]

[1] Dipartimento di Fisica, Sapienza Università di Roma, P. le A. Moro 2, 00185 Rome, Italy
[2] INFN Sezione di Roma, Sapienza Università di Roma, P.le A. Moro 2, 00185 Rome, Italy
[3] Dipartimento di Fisica "Enrico Fermi", Università di Pisa, Largo Bruno Pontecorvo 3, 56127 Pisa, Italy
[4] INFN Sezione di Pisa, Largo Bruno Pontecorvo 3, 56127 Pisa, Italy
[5] Instituto de Física, Universidad Nacional Autónoma de México, A.P. 20-364, 01000 Ciudad de México, Mexico
[6] INFN Sezione di Ferrara, Via Saragat 1, 44122 Ferrara, Italy
[7] University Grenoble Alpes, CNRS, Grenoble INP, Institut Néel, 38000 Grenoble, France
[8] INFN Laboratori Nazionali del Gran Sasso, Via Sommarive 14, 67100 Assergi, AQ, Italy
[9] Institute for Data Processing and Electronics, Karlsruhe Institute of Technology, Hermann-von-Helmholtz-Platz 1, 76344 Eggenstein-Leopoldshafen, Germany
[10] INFN-TIFPA, Via Sommarive 14, 38123 Povo (Trento), Italy
[11] Dipartimento di Fisica e Scienze della Terra, Università di Ferrara, Via Saragat 1, 44100 Ferrara, Italy
[12] Gran Sasso Science Institute, Viale Francesco Crispi, 7 Rectorate, Via Michele Iacobucci, 2, 67100 L'Aquila, AQ, Italy

Received: 21 October 2025 / Accepted: 6 March 2026

**Abstract** BULLKID is a cryogenic, solid-state detector designed for direct searches of particle Dark Matter candidates, with mass $\lesssim$ 1 GeV/c$^2$, and coherent neutrino-nucleus scattering. It is based on an array of dice carved in 5 mm thick silicon crystal, sensed by phonon-mediated Kinetic Inductance Detectors. In previous works, the array was calibrated with bursts of optical photons, which are absorbed in the first hundreds nanometers of the dice and give rise to surface events. In this work, we present the reconstruction of bulk events through the 59.5 keV $\gamma$-ray generated by an $^{241}$Am source, which emulates more closely the interaction of Dark Matter and neutrinos. The peak resolution is 5% ($\sigma$) and its position is shifted by less than 10% with respect to the optical calibration. We observe that the resolution is further improved by a factor 2 combining the signal from neighboring dice. These results confirm the performance of the detector in view of the physics goals of the BULLKID-DM experiment for dark matter search.

## 1 Introduction

BULLKID [1–3] is a monolithic array of dice of 5.4 × 5.4 × 5.0 mm$^3$ carved in a silicon crystal and sensed by phonon-mediated cryogenic Kinetic Inductance Detectors (KIDs) [4–10]. It is designed for the detection of sub-keV energy depositions from particle interactions within the crystal, making it suitable for direct Dark-Matter (DM) searches [11–16] and coherent elastic neutrino-nucleus scattering (CE$\nu$NS) [17–19] experiments. One of the main challenges in this low-energy regime is the energy calibration, which is currently performed with a method based on bursts of 400 nm optical photons [9,20]. These photons produce electron recoils near the surface of the crystal, unlike DM or CE$\nu$NS interactions, which produce nuclear recoils uniformly throughout the crystal volume (bulk events). Validating this optical calibration method, using bulk events from particle interactions, is thus of critical importance to confirm the detector's performance under realistic conditions. In this work, we present such a validation by reconstructing the 59.5 keV $\gamma$-ray peak from a radioactive $^{241}$Am source using a BULLKID detector. These $\gamma$-rays are energetic enough to penetrate the entire crystal, with an absorption length $\lambda_{abs}$ of 13 mm [21] – defined as the distance over which a photon has a $1 - 1/e$ probability of being absorbed. In contrast, optical photons are almost entirely absorbed at the surface, with an absorption length of approximately 180 nm [22]. The measurement presented in this work checks the reliability of the optical calibration technique and allows the investigation for different responses between bulk and surface events.

[a] e-mail: matteo.folcarelli@uniroma1.it (corresponding author)

## 2 Experimental setup

The BULLKID detector exploited in this measurement consists in a monolithic silicon wafer, 3 inches in diameter and 5 mm thick, carved in 60 dice of 0.34 g each (Fig. 1 Top). The carvings leave intact a 0.5 mm thick common disk that acts as both the holding structure of the array and as substrate for the aluminum lithography of the KIDs [1]. Figure 1 (center) shows a picture of the detector lithography and a map of its resonators. We identify a total number of 57 working KIDs out of 60, due to fabrication issues, with the median value of the quality factor $Q$ distribution of 120 k. For operation, the wafer is installed in a copper holder by clamping the crystal in the peripheral region of the wafer (Fig. 1 Top) and placed inside aluminum and Cryophy®pots for shielding from thermal radiation and external magnetic fields. The system is anchored to the coldest point of a dry $^3$He/$^4$He dilution refrigerator with a base temperature of 20 mK. In addition, the cryostat is surrounded by a lead castle to shield it from external radioactivity.

The energy of a particle interacting in a die is converted to athermal phonons (Fig. 2) that, in part, are absorbed by the KID or by non-sensitive elements, e.g. the feedline. The absorption in the KID leads to a phase shift in the resonator's transfer function which, for small phase variations, is proportional to the deposited energy [24]:

$$\Delta\phi \propto Q \cdot \Delta E \tag{1}$$

As the energy increases, the phase response saturates, introducing a non-linearity in the detector's response function. The rest of the phonons leak through the common disk in nearby dice and are absorbed by the respective KIDs. In a previous work on a different BULLKID detector [1], with respect to the energy measured in the originating die, an average of (14 ± 3)% of the energy was observed in each of the first neighbors along the vertical and horizontal directions, and (5 ± 1)% in the diagonal ones. This effect is exploited to determine if an event originates in the die corresponding to the KID considered [2].

The energy calibration is performed by shining controlled photon bursts to the KID 46, 47 and 49 on the opposite side of the lithography, produced by a room temperature 400 nm LED lamp and delivered through optical fibers. The number of available fibers limits the calibration to 3 KIDs. The $^{241}$Am source has an activity of 150 Bq and decays spontaneously to $^{237}$Np by emitting a 59.5 keV $\gamma$-ray. It is fixed on the head of a screw anchored 5 cm below the aluminum pot (Fig. 1 Bottom).

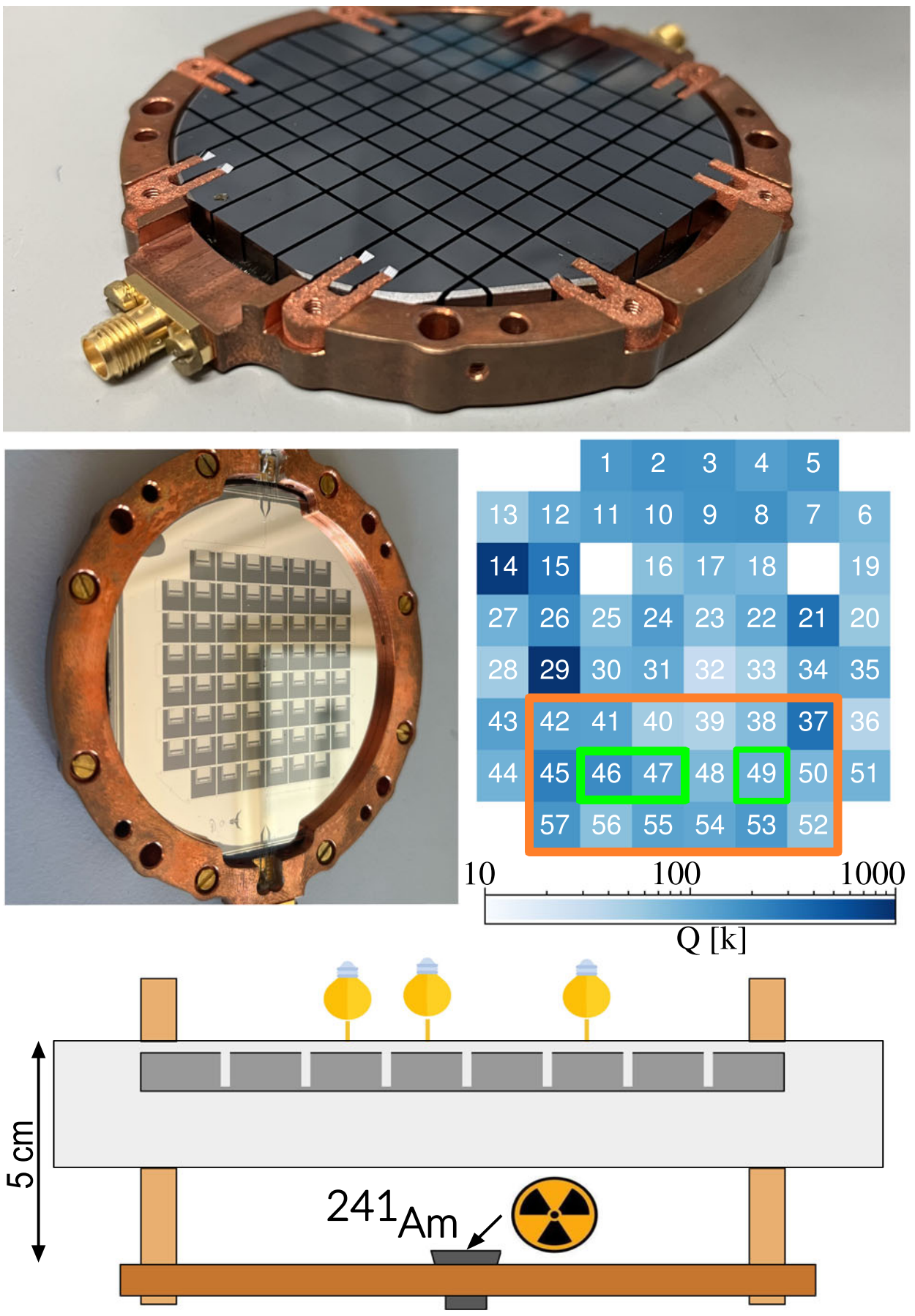

**Fig. 1** Top: silicon wafer of BULLKID. It is made of 60 silicon dice of 5.4×5.4×5 mm$^3$ and 0.34 g each. It is installed in a copper holder for the mechanical stacking and the thermalization at cryogenic temperatures; center: picture of the wafer (from the side of the lithography) and map of its working KIDs. The resonators inside the orange continuous line represent the acquired KIDs when the main resonators 46, 47, 49, in the green boxes, trigger. The empty cells represent missing KIDs while the colorscale is representative of the total quality factor of each resonator. Bottom: scheme (not in scale) of the aluminum pot (light grey) that hosts the BULLKID detector (dark grey) and of the screw with the Americium source on its head. On the top of the aluminum pot, a sketch of the optical fibers used for the illumination of the main KIDs

During the data-taking, the KIDs in the orange contour of Fig. 1 (center) are acquired simultaneously when one of the resonators 46, 47, 49, dubbed as "main KIDs", triggers. The strategy of the acquisition has the aim of fiducializing the active mass, discarding all the events in which the interaction did not take place in a main die. The incoming data stream is triggered online, selecting all the signals with an amplitude 30 times greater than the standard deviation of the noise.

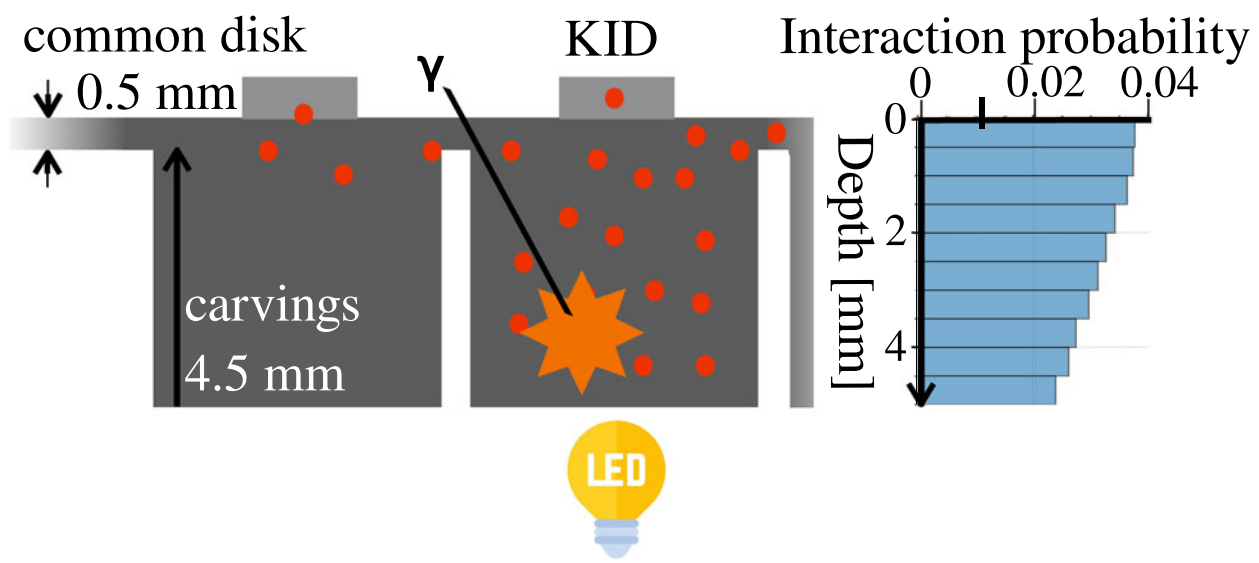

**Fig. 2** Schematic representation of a 59.5 keV $\gamma$-ray emitted from a radioactive $^{241}$Am source interacting within BULLKID. The plot on the right shows the interaction probability, evaluated through a GEANT4 [23] simulation, of such $\gamma$-rays as a function of the depth in the silicon crystal. The depth origin is referenced from the top of the common disk of the structure, as the radioactive source is positioned on the lithography side. Athermal phonons (red dots) generated by the $\gamma$-ray interaction are predominantly collected by the KID located on the corresponding die. A fraction of these phonons leaks through the common disk into adjacent dice, inducing signals in their respective KIDs. For reference, the light bulb represents the optical fiber, used for the optical calibration, that shines the side of the die opposite to the lithography

**Table 1** Calibration constant of the main KIDs evaluated through the combination of the LED calibrations at the beginning and at the end of the data-acquisition. In the last column it is shown the $b$ parameter of the quadratic relation (Eq. 2) that models the non-linearity of the KID

| KID | LED calib. constant [eV/mrad] | b [$10^{-4}$ mrad$^{-1}$] |
|---|---|---|
| 46 | $47.2 \pm 3.1$ | − 1.2 |
| 47 | $77.5 \pm 6.0$ | − 1.4 |
| 49 | $90.1 \pm 6.5$ | − 1.4 |

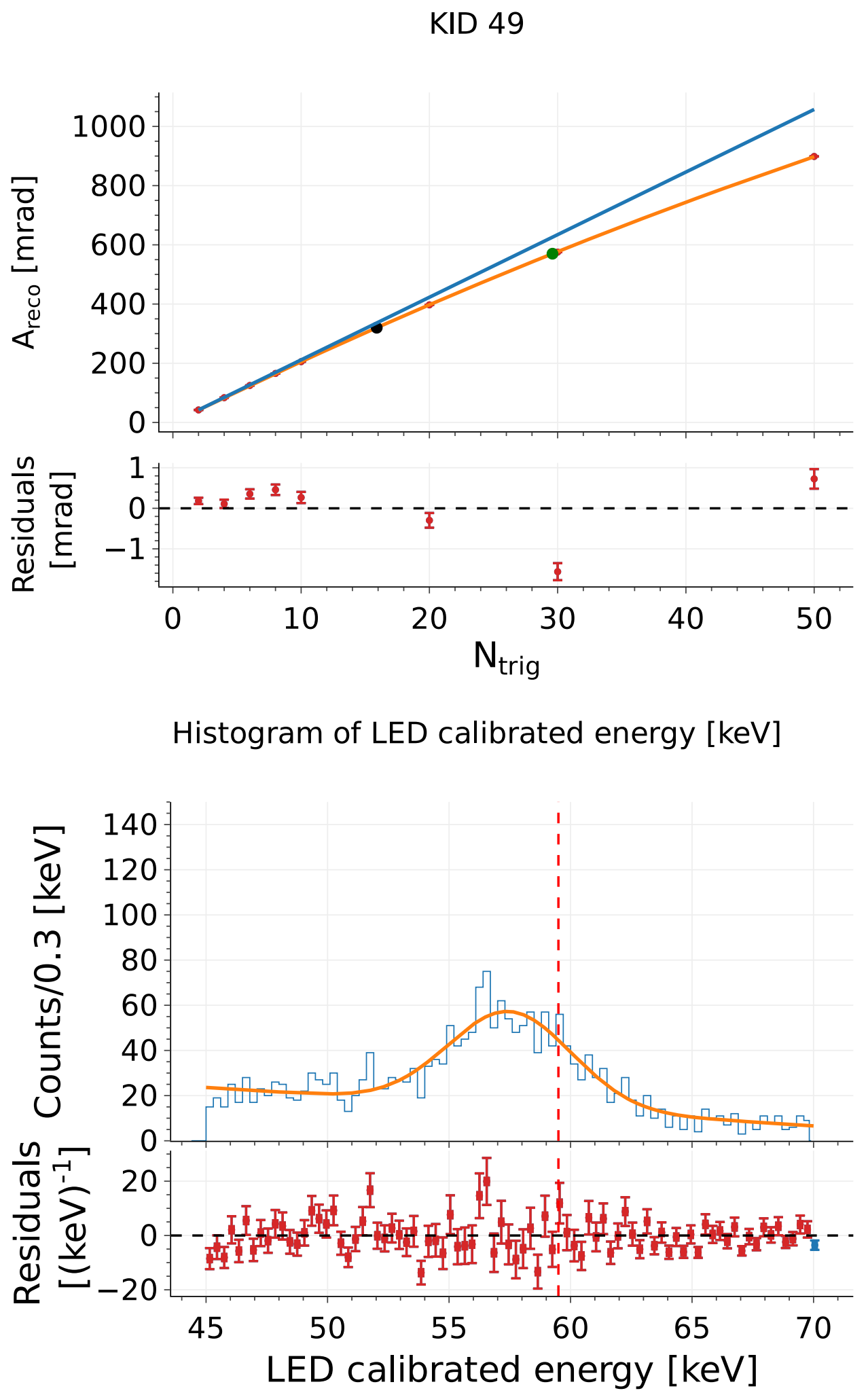

**Fig. 3** Top: dependence of the reconstructed amplitude $A_{reco}$ to the number of triggers $N_{trig}$ for the main KID 49. The orange curve corresponds to the fit of of the data with the model described in Eq. 2 and the blue line represents the response function after the correction for the non-linearity by means of Eq. 3. The black and green dots are the reconstructed amplitudes of the control LED and of the Americium peak; bottom: energy spectrum of the same KID in the range [45, 70] keV calibrated with optical photons. The Americium peak is identified and fitted with a Gaussian model superimposed on the graph. The red-dashed vertical line represents the 59.5 keV true energy

The entire measurement consists of 28 h of data-acquisition performed in sets of 1 h separated by intervals of 10 min. During these intervals, an acquisition of noise streams and LED pulses is performed in order to monitor the detector stability. No significant change of the detector response have been detected over time.

## 3 Data analysis and results

The amplitude of the triggered signals is estimated offline as the distance between the maximum and the baseline of the waveform. An optical calibration [9], at the beginning and at the end of the data-taking, allows the evaluation of the calibration constant of the main KIDs giving the results presented in Table 1. Throughout the data-taking period, LED pulses are periodically sent to the main KIDs to monitor the stability of their response. The LED lamp is controlled via an external trigger that we fire at increasing rate in order to increase the number of photons in a burst. The dependence of the reconstructed amplitude $A_{reco}$ to the number of these external triggers $N_{trig}$ is a measure of the detector's linearity and we model it through the parabolic relation

$$A_{reco} = a \cdot N_{trig} \cdot (1 + a \cdot b \cdot N_{trig}). \tag{2}$$

The parameter $b$, reported in Table 1, represents the non linearity of the response function inferred with the fit in Fig. 3 (top). The black and the green dots, at $N_{trig} \sim 15$ and 30, are the reconstructed amplitudes of the control LED and of the Americium peak, corresponding respectively to 30 and 59.5 keV. The linearized amplitude $A_{lin} = a \cdot N_{trig}$ is evaluated

**Table 2** Results of the Gaussian fits to the Americium peaks for the three main KIDs. The columns report the Mean and the standard deviation $\sigma$ for the single die energy and the standard deviation for the calorimetric energy $\sigma_{cal}$. The last column shows the relative deviation of the measured peak position from the nominal value

| KID | Mean [keV] | $\sigma$ [keV] | $\sigma_{cal}$ [keV] | Mean - nominal |
|---|---|---|---|---|
| 46 | $56.2 \pm 0.1$ | $2.2 \pm 0.1$ | $1.1 \pm 0.1$ | −6% |
| 47 | $55.5 \pm 0.2$ | $2.4 \pm 0.2$ | $1.1 \pm 0.1$ | −7% |
| 49 | $57.2 \pm 0.2$ | $2.9 \pm 0.2$ | $1.3 \pm 0.1$ | −4% |

by inverting Eq. 2.

$$A_{\text{lin}} = \frac{-1 + \sqrt{1 + 4 \cdot b \cdot A_{\text{reco}}}}{2b}. \tag{3}$$

In the following, the "LED calibrated energy" denotes the linearized amplitude $A_{\text{lin}}$ calibrated with the calibration constant of the corresponding main KID.

After applying analysis cuts based on pulse shape variables and the neighbor algorithm [2], which select only interactions occurring in the main dice, we reconstruct the energy spectra of the main KIDs. Fig. 3 (bottom) shows such energy spectrum for KID 49 in the energy range [45, 70] keV where a peak is clearly reconstructed and identified as the Americium peak. A fit with a linear decreasing background and a Gaussian model of the peak is superimposed to the energy spectrum. For the main KIDs the inferred parameters of the Americium peak are summarized in Table 2. The energy resolution amounts to around 5% ($\sigma$) while the mean is reconstructed with a deficit of a few % with respect to the nominal energy, pointing to a systematics of the LED calibration technique. Nevertheless such deficit is smaller than 10%, validating the optical calibration of the BULLKID detector units to such level of accuracy. The physical origin of this discrepancy might be due to the methodology of the LED calibration, which is based on the Poisson process of the absorbed photons [9] which in turns are converted to phonons. The poissonian distribution of the generated phonons has not been taken into account so far and it might be the cause of this discrepancy. This study requires additional measurements, which will be the subject of future work.

## 4 Neighbor analysis for bulk events

Effects related to the interaction in the bulk position of Americium can be investigated by studying the partition between main and neighboring dice. This is evaluated as the ratio of the amplitudes of the two dice ($A$ for the main KIDs and $A_n$ for the neighbors), corrected for their corresponding quality factors from Eq. 1:

$$R_n = \frac{A_n}{A} \cdot \frac{Q}{Q_n} \tag{4}$$

For reference, Fig. 4 (top) shows the energy ratio of all the surrounding neighbors of KID 49 as a function of the LED calibrated energy. The position of the subplot corresponds to the neighbor considered; the central one shows the energy spectrum of KID 49. For all the main KIDs, we measure an average energy ratio of $(17 \pm 1)\%$ in the vertical and horizontal directions and $(10 \pm 1)\%$ in the diagonal one. With respect to a previous measurement [1], we observe an increase that we associate to a possible variation of the groove depth or to a reduction of the passive metal on the lithography. The bottom dice measure a larger ratio (about 25%) due to the position of the resonator, centered in the upper part of the dice. In Fig. 4 (top) two clusters are clearly visible: the control LED around 30 keV and the Americium peak around 57 keV. The inset in each subplot is a close up on the Americium cluster that presents an anti-correlation: Americium events are mono-energetic, hence the more phonons leak in the neighbors the less are absorbed by the main KID.

Additionally, the energy ratios between neighboring dice for Americium events correlate. Figure 4 (bottom) shows the energy ratios for events in the main KID 49 in the energy range [45, 70] keV, between the dice: left-right, top-bottom and diagonal corners.[1] Such correlation suggest that the effects we observe for Americium events are not related to the position of the events on the plane parallel to the lithography, but to the depth in the silicon dice. In particular, a possible origin of the lower energy reconstructed with respect to the LED calibration could be the greater phonon leakage produced by events closer to the common disk. LED events, that all happen on the surface of a die, and have an estimated light spot size of approximately 1 mm, do not show any sign of correlation between neighboring dice.

Following the model of the phonon leakage just exposed, the sum of the amplitudes of the KIDs (main + neighbors) should be able to compensate for the anti-correlation, improving the resolution of the $^{241}$Am peak. In this regard, we define the "calorimetric" amplitude:

$$A_{\text{cal}} = A/Q + \sum_{\text{neigh.}} A_n/Q_n \tag{5}$$

[1] The same positive correlation is observed for all neighbor pairs and for all the main KIDs.

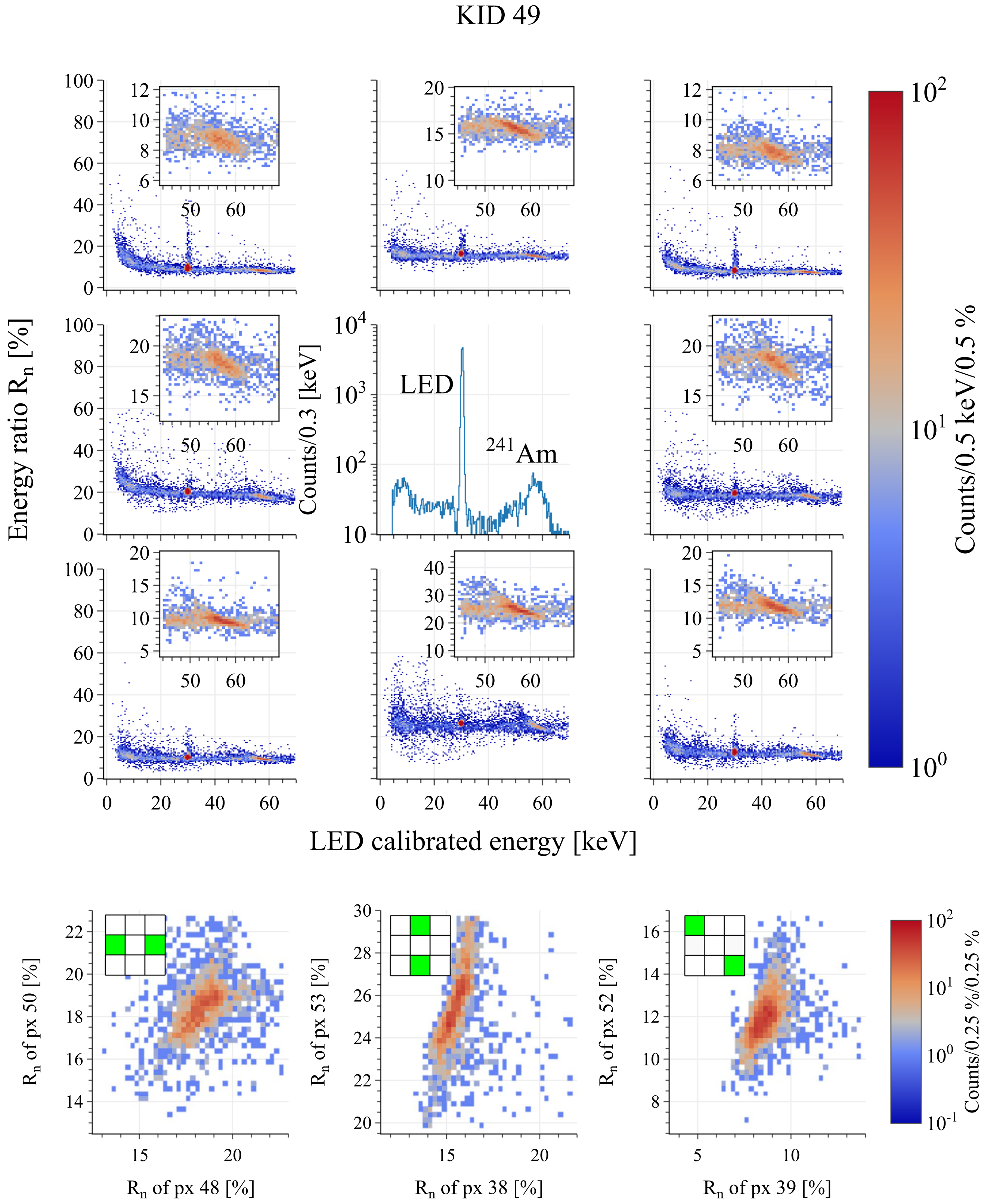

**Fig. 4** Top: energy ratio $R_n$, as defined in Eq. 4, for the neighbor dice of the KID 49 as a function of the LED calibrated energy in the main die. The position of the single plots corresponds to the die to which the ratio is evaluated. In the center of the plot, the energy spectrum of KID 49. The insets in each cell is a close up on the region of Americium events ([45, 70] keV) in order to point out the anti-correlation that we measure for each neighbour; bottom: energy ratio $R_n$ between three couples of neighbors of KID 49: left/right, top/bottom and diagonal corners. Only events in the energy range [45, 70] keV in the main die are considered. We observe a positive correlation in all the three cases and the same is for all the possible neighbors pairs and for all the main KIDs

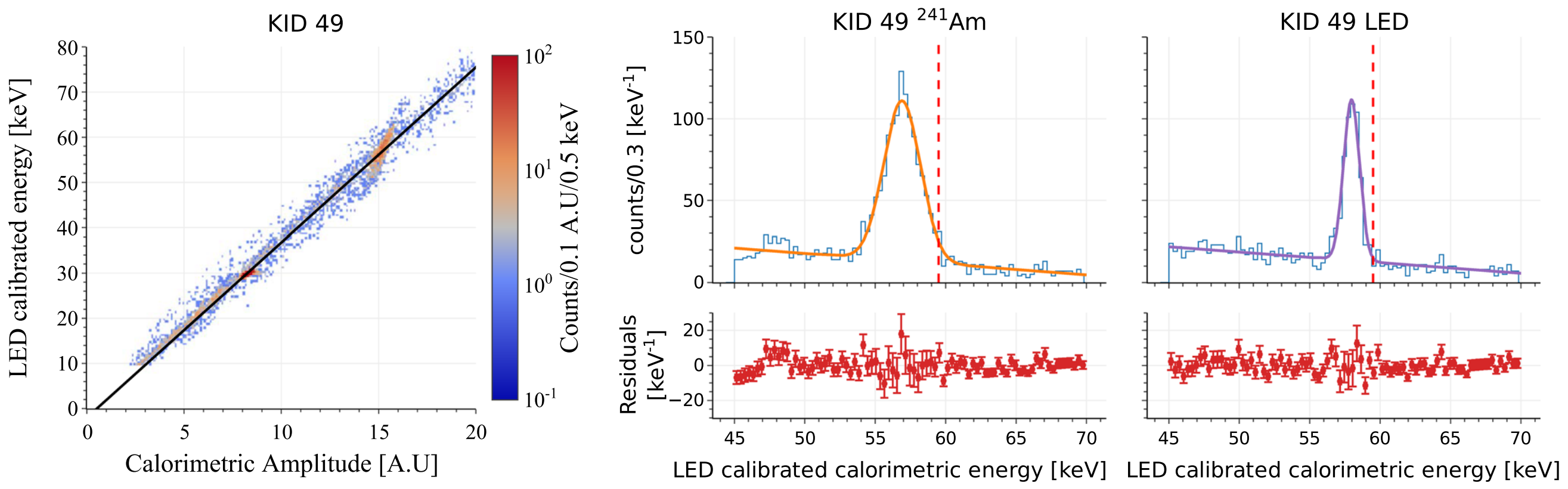

**Fig. 5** Left: linear relation between the calorimetric and the LED calibrated energy for the main KID 49 in the energy range [10, 80] keV. The black fit superimposed gives the calibration parameter to convert the former amplitude in keV; center: energy spectrum of KID 49 in the range [45, 70] keV with the calorimetric amplitude defined in Eq. 5. The Americium peak is identified and fitted with a Gaussian model superimposed on the graph. The red-dashed vertical line represents the 59.5 keV nominal energy; right: energy spectrum of KID 49 in the range [45, 70] keV with the calorimetric amplitude defined in Eq. 5 during an optical calibration. The LED peak is identified and fitted with a Gaussian model superimposed on the graph. The red-dashed vertical line represents the 59.5 keV nominal energy of Americium

where the neighbors included in the sum are limited to those surrounding the main dice, as they exhibit the highest energy ratio. For increasing energy deposited in the main die, more signal leaks into neighboring dice. For this reason, we expect a linear relation between the new calorimetric amplitude $A_{\mathrm{cal}}$ – which includes this leakage – and the (LED-calibrated) energy in the main die, as shown for KID 49 in Fig. 5 (left). We exploit this linear correlation by performing a fit, for all the main KIDs, using all the events in the energy range [10, 80] keV. The result of this fits provides a calibration function that maps the calorimetric amplitude to energy, enabling the reconstruction of the new energy spectrum of the Americium peak.

In Fig. 5 (center) such peak is fitted with a Gaussian model and the parameters are listed in Table 2. No difference (within the errors of the fit) is found between the single die and the calorimetric mean value of the Americium peak. An improvement of the resolution by a factor 2 is measured, reaching the value of 2% ($\sigma$). In Fig. 5 (right) it is represented one of the LED acquisitions, used for the optical calibration, in the same energy range of Americium. We evaluate a resolution of 1% ($\sigma$) that is dominated by the photo-statistics of multiple photon bursts. This suggests that the calorimetric energy does not fully correct the dependence on the interaction point or other effects of energy loss. Moreover, LED events are not only localized in depth but also along the plane parallel to the lithography. In contrast, Americium events are uniformly distributed across the entire detector surface.

## 5 Conclusions

We performed the calibration of the bulk events induced by the 59.5 keV $\gamma$-rays of a $^{241}$Am radioactive source with a silicon BULLKID detector. Adopting a calibration based on optical photons, the mean of all the reconstructed peaks presents a deficit with respect to the true energy, suggesting a systematic bias in the calibration technique. However, such deficit is less than 10%, validating the optical calibration, of the BULLKID detector, to such level of accuracy. We observe an anti-correlation in Americium events between the energy deposited in the main die and in neighboring dice pointing to positional effects related to the depth of the interaction point. By summing the energies of nearby dice, we are able to correct for such anti-correlation improving the peak resolution from 5 to 2% ($\sigma$).

**Acknowledgements** We acknowledge the support of the Doctoral School *"Karlsruhe School of Elementary and Astroparticle Physics: Science and Technology"*, the SECIHTI Project No. CBF-2025-I-1589 and DGAPA UNAM Grants No. PAPIIT IN105923 and IN102326. This work was further supported by the INFN, Sapienza University of Rome and co-funded by the European Union (ERC, DANAE, 101087663). Views and opinions expressed are however those of the author(s) only and do not necessarily reflect those of the European Union or the European Research Council. Neither the European Union nor the granting authority can be held responsible for them. We thank A. Girardi and M. Iannone of the INFN Sezione di Roma for technical support. We thank R. Mirabelli of the Department of SBAI at Sapienza University for useful discussions.

**Data Availability Statement** Data will be made available on reasonable request. [Author's comment: The datasets generated during and/or analysed during the current study are available from the corresponding author on reasonable request.]

**Code Availability Statement** Code/software will be made available on reasonable request. [Author's comment: The code/software generated during and/or analysed during the current study is available from the corresponding author on reasonable request.]

Funded by SCOAP[3].